
\magnification=\magstep1
\voffset= 0.0 truein
\hoffset=0.0 truein
\vsize=9.0 truein
\hsize=6.5 truein
\parskip=5truept plus2truept
\baselineskip 0.25 truein
\raggedbottom


\def\HII{\hbox{H\hskip 2trueptII }}
\def\degns{\ifmmode^\circ\else$^\circ$\fi}
\def\deg{\ifmmode^\circ\else$^\circ$\fi\ }
\def\solar{\ifmmode_{\mathord\odot}\else$_{\mathord\odot}$\fi}
\def\arcs{\char'175\ }
\def\arcm{\char'023\ }

\def\eg{e.g.,\ }

\def\etal{et~al.\ }

%
%
\newcount\figureno  \newcount\tableno
\figureno=0         \tableno=0
\newbox\tablebox    \newdimen\tablewidth
\def\section#1 #2\par{\vskip .5cm \goodbreak
    \centerline{{\sr#1.\ #2}}\nobreak\medskip}
\def\subsection#1 #2\par{\vskip .3cm
    \ifdim\lastskip=\medskipamount \nobreak \else\goodbreak\fi
    \centerline{{\it#1\/})\ {\it #2}}\nobreak\medskip}
\def\subsubsection#1 #2\par{\vskip .3cm
    \ifdim\lastskip=\medskipamount \nobreak \else\goodbreak\fi
    \centerline{{\er#1)}\ {\si #2}}\nobreak\medskip}
\def\today{\number\day\space \ifcase\month\or January\or February\or
    March\or April\or May\or June\or July\or August\or
    September\or October\or November\or December\fi \space\number\year}
\def\deg{\ifmmode ^{\circ}
         \else $^{\circ}$\fi}
\def\pdeg{\ifmmode
           $\setbox0=\hbox{$^{\circ}$}\rlap{\hskip.11\wd0 .}$^{\circ}
     \else \setbox0=\hbox{$^{\circ}$}\rlap{\hskip.11\wd0 .}$^{\circ}$\fi}
\def\arcs{\ifmmode {''}\else $''$\fi}
\def\arcm{\ifmmode {'}\else $'$\fi}
\newdimen\sa  \newdimen\sb
\def\parcs{\sa=.07em \sb=.03em
     \ifmmode $\rlap{.}$^{\scriptscriptstyle\prime\kern -\sb\prime}$\kern -\sa$
     \else \rlap{.}$^{\scriptscriptstyle\prime\kern -\sb\prime}$\kern -\sa\fi}
\def\parcm{\sa=.08em \sb=.03em
     \ifmmode $\rlap{.}\kern\sa$^{\scriptscriptstyle\prime}$\kern-\sb$
     \else \rlap{.}\kern\sa$^{\scriptscriptstyle\prime}$\kern-\sb\fi}

\def\exp#1 {\ifmmode \times 10^{#1}
            \else $\times 10^{#1}$\fi}
\def\page{\vfill\eject}
\def\leaderfil{\leaders\hbox to 5pt{\hss.\hss}\hfil}
%
\def\jref#1 #2 #3 #4 {{\par\noindent \hangindent=3em \hangafter=1
      \advance \rightskip by 5em #1, {\it#2}, {\bf#3}, #4.\par}}
\def\ref#1{{\par\noindent \hangindent=3em \hangafter=1
      \advance \rightskip by 5em #1.\par}}
%
%
%
%
%
\newif\iftform      \newif\ifmform      \newif\ifdraft
\newcount\toes      \toes=0
\newif\ifcapsopen   \newif\iftabsopen
\newwrite\caps      
\newwrite\tabs      
\def\raggedright{\rightskip=0pt plus6em
    \spaceskip=.3333em \xspaceskip=.5em }

\def\datereceived#1\par{\vskip 2pt
    \centerline{{\it Received #1; accepted\rcvrule}}
    \ifmform \runtitle \fi}
\def\rcvrule{\hskip 4pt\vrule height 0pt width 90pt depth 1pt}
\def\runtitle{\vskip 1in
    \centerline{{\it Running Title:}\quad\title}}
\def\abstract "#1" "#2"{\ifmform \page\fi
   {\iftform \advance\leftskip by .25in \advance\rightskip by .25in\fi
    \vskip .7cm \centerline{ABSTRACT} \vskip .3cm
    #1
    \vskip 1pt
    {\setbox3=\hbox{\it Subject headings:\quad}
     \hangindent=\wd3  \hangafter=1  \noindent
     \box3 #2 \par}}
    \ifmform \page\fi}
\def\ajabstract "#1"{\ifmform \page\fi
   {\iftform \advance\leftskip by .25in \advance\rightskip by .25in\fi
    \vskip .7cm \centerline{ABSTRACT} \vskip .3cm
    #1}
    \ifmform \page\fi}
%
%

\def\figure #1 {\advance\figureno by 1
        \ifmform \ifcapsopen\relax
                 \else\immediate\openout\caps=figures.jnk\capsopentrue\fi
            \vskip .7cm
            \centerline{FIGURE \number\figureno}
            \vskip .7cm
            \immediate\write\caps{\string\efigure}\copytoblankline
        \fi%
        \iftform \eightpoint
           \ifdim#1>7.995in \pageinsert\vskip 0pt plus 16000fill
               F{\sc IG} \number\figureno .---\readtoblankline
           \fi
           \ifdim#1<7.995in \midinsert\vskip #1
              F{\sc IG} \number\figureno .---\readtoblankline
           \fi
        \fi}
\def\etable #1\par{\advance\tableno by 1
            \ifmform \vfill{\eightpoint\input #1 }\page \fi}
\def\efigure #1\par{\advance\figureno by 1
               \ifmform F{\sr IG} \number\figureno .---#1\par\fi}
%
%
\def\readtoblankline{\obeylines\readcap}
{\obeylines  \gdef\readcap#1
  {\def\next{#1}%
   \ifx\next\empty\let\next=\endread %
   \else\next\space \let\next=\readcap\fi\next}}
\def\endread{\vfil\endinsert\tenpoint}
\def\copytoblankline{\begingroup\setupcopy\copycap}
{\obeylines  \gdef\copycap#1
  {\def\next{#1}%
   \ifx\next\empty\let\next=\endcopy %
   \else\immediate\write\caps{\next} \let\next=\copycap\fi\next}}
\def\endcopy{\endgroup\immediate\write\caps{\string\par}}
\chardef\other=12
\def\setupcopy{\def\do##1{\catcode`##1=\other}\dospecials
               \catcode`\|=\other \obeylines}

\def\foot#1{\advance\toes by 1
     \ifmform $^{\number\toes}$\vadjust{
        \kern\baselineskip\hrule\vskip 7pt\noindent$^{\number\toes}$ #1
        \vskip 7pt\hrule\vskip\baselineskip}
     \else{\eightpoint

\unskip\unskip\unskip\unskip\unskip\footnote{$^{\number\toes}$}{#1}}\fi}
\def\specialfoot#1#2{
     \ifmform $^{#1}$\vadjust{
        \kern\baselineskip\hrule\vskip 7pt\noindent$^{#1}$ #2
        \vskip 7pt\hrule\vskip\baselineskip}
     \else{\eightpoint
        \unskip\unskip\unskip\unskip\unskip\footnote{$^{#1}$}{#2}}\fi}

%

\def\expand{
\advance \hsize by +0.1 true in \advance \hoffset by -0.05 true in
\advance \vsize by +0.1 true in \advance \voffset by -0.05 true in}
\def\contract{
\advance \hsize by -0.1 true in \advance \hoffset by +0.05 true in
\advance \vsize by -0.1 true in \advance \voffset by +0.05 true in}

\def\clock{\count0=\time \divide\count0 by 60
    \count1=\count0 \multiply\count1 by -60 \advance\count1 by \time
    \number\count0:\ifnum\count1<10{0\number\count1}\else\number\count1\fi}
\def\deg{\ifmmode^\circ\else$^\circ$\fi}
\def\solar{\ifmmode_{\mathord\odot}\else$_{\mathord\odot}$\fi}
\def\er{\relax} \def\sr{\relax}

\newcount\eqnum
\def\nexteq{\global\advance\eqnum by1 \eqno(\number\eqnum)}
\def\lasteq#1{\if)#1[\number\eqnum]\else(\number\eqnum)\fi#1}
\def\preveq#1#2{{\advance\eqnum by-#1
    \if)#2[\number\eqnum]\else(\number\eqnum)\fi}#2}

\def\page{\vfill\eject}
\def\etal{\it et al. \rm }


\def\tableheight{\vrule width 0pt height 8.5pt depth 3.5pt}
{\catcode`|=\active \catcode`&=\active
    \gdef\tabledelim{\catcode`|=\active \let|=\vbar
                     \catcode`&=\active \let&=\nobar} }
\def\table{\begingroup
    \def\twidth{\hsize}
    \def\tablewidth##1{\def\twidth{##1}}
    \def\defaultheight{\vrule width 0pt height 8.5pt depth 3.5pt}
    \def\heightdepth##1{\dimen0=##1
        \ifdim\dimen0>5pt
            \divide\dimen0 by 2 \advance\dimen0 by 2.5pt
            \dimen1=\dimen0 \advance\dimen1 by -5pt
            \vrule width 0pt height \the\dimen0  depth \the\dimen1
        \else  \divide\dimen0 by 2
            \vrule width 0pt height \the\dimen0  depth \the\dimen0 \fi}
    \def\spacing##1{\def\defaultheight{\heightdepth{##1}}}
    \def\nextheight##1{\noalign{\gdef\tableheight{\heightdepth{##1}}}}
    \def\end{\cr\noalign{\gdef\tableheight{\defaultheight}}}
    \def\zerowidth##1{\omit\hidewidth ##1 \hidewidth}
    \def\hline{\noalign{\hrule}}
    \def\skip##1{\noalign{\vskip##1}}
    \def\bskip##1{\noalign{\hbox to \twidth{\vrule height##1 depth 0pt \hfil
        \vrule height##1 depth 0pt}}}
    \def\header##1{\noalign{\hbox to \twidth{\hfil ##1 \unskip\hfil}}}
    \def\bheader##1{\noalign{\hbox to \twidth{\vrule\hfil ##1
        \unskip\hfil\vrule}}}
    \def\spanloop{\span\omit \advance\mscount by -1}
    \def\extend##1##2{\omit
        \mscount=##1 \multiply\mscount by 2 \advance\mscount by -1
        \loop\ifnum\mscount>1 \spanloop\repeat \ \hfil ##2 \unskip\hfil}
    \def\vbar{&\vrule&}
    \def\nobar{&&}
    \def\hdash##1{ \noalign{ \relax \gdef\tableheight{\heightdepth{0pt}}
        \toks0={} \count0=1 \count1=0 \putout##1\end
        \toks0=\expandafter{\the\toks0 &\end} \xdef\piggy{\the\toks0} }
        \piggy}
    \let\e=\expandafter
    \def\putspace{\ifnum\count0>1 \advance\count0 by -1
        \toks0=\e\e\e{\the\e\toks0\e&\e\multispan\e{\the\count0}\hfill}
        \fi \count0=0 }
    \def\putrule{\ifnum\count1>0 \advance\count1 by 1

\toks0=\e\e\e{\the\e\toks0\e&\e\multispan\e{\the\count1}\leaders\hrule\hfill}
        \fi \count1=0 }
    \def\putout##1{\ifx##1\end \putspace \putrule \let\next=\relax
        \else \let\next=\putout
            \ifx##1- \advance\count1 by 2 \putspace
            \else    \advance\count0 by 2 \putrule \fi \fi \next}   }
\def\tablespec#1{
    \def\vdimens{\noexpand\tableheight}
    \def\tabby{\tabskip=0pt plus100pt minus100pt}
    \def\r{&################\tabby&\hfil################\unskip}
    \def\c{&################\tabby&\hfil################\unskip\hfil}
    \def\l{&################\tabby&################\unskip\hfil}
    \edef\templ{\noexpand\vdimens ########\unskip  #1
         \unskip&########\tabskip=0pt&########\cr}
    \tabledelim
    \edef\body##1{ \vbox{
        \tabskip=0pt \offinterlineskip
        \halign to \twidth {\templ ##1}}} }
\def\ref{\noindent \hangindent=3em \hangafter=1}
\def\solar{\ifmmode_{\mathord\odot}\else$_{\mathord\odot}$\ \fi}

\font\tenrm=cmr10
\def\sb{\ifmmode{\;{\rm mag}\;{\rm arcsec}^{-2}}\else{~mag~arcsec$^{-2}$}\fi}
\def\HII{\hbox{H\hskip 2trueptII }}
\def\HI{\hbox{H\hskip 2trueptI }}
\def\pop3{Pop.~III}

\def\eg{e.g.,\ }

\def\cf{cf.\ }
\def\etal{et al.\ }
\def\r23{R$_{23}$}
\def\o32{O$_{32}$}
\def\R23{$R_{23}$}
\def\O32{$O_{32}$}

\def\caption{\baselineskip 8pt \lineskip 2pt \parskip 8pt plus 1pt}
\def\gtsima{$\; \buildrel > \over \sim \;$}
\def\ltsima{$\; \buildrel < \over \sim \;$}
\def\prosima{$\; \buildrel \propto \over \sim \;$}
\def\gsim{\lower.5ex\hbox{\gtsima}}
\def\lsim{\lower.5ex\hbox{\ltsima}}
\def\simgt{\lower.5ex\hbox{\gtsima}}
\def\simlt{\lower.5ex\hbox{\ltsima}}
\def\simpr{\lower.5ex\hbox{\prosima}}
\newcount\tablecounter
     \tablecounter=0
     \long\def\fudgetable#1{%
     \iffirsttable\global\doinglistoftabpagetrue\fi
     \global\firsttabletrue
          \greatbreak\vskip .005in plus.005in minus.001in\greatbreak
          \global\advance\tablecounter by 1
          \centerpar{\bf Table \number\chaptercounter.%
                    \number\tablecounter\ --- #1}%
          \nobreak\vskip .005in minus.001in\RSreset\nobreak
          \listentry{1}{\number\chaptercounter.\number\tablecounter}{#1}{1}}
\newcount\eqnum
\eqnum=0 
\def\numeqn {\global\advance\eqnum by 1
             \eqno(\number\chaptercounter.\number\eqnum)}
\newcount\apeqnum
\apeqnum=0 
\def\apnumeqn#1{\global\advance\apeqnum by 1
             \eqno(#1.\number\apeqnum)}

\font\tenrm=cmr7
\baselineskip=12pt

\centerline{\bf THE MORPHOLOGY OF LOW SURFACE BRIGHTNESS DISK GALAXIES}
\vskip 1.0in
\centerline{S{\tenrm TACY} S.\ M{\tenrm C}G{\tenrm AUGH}}
\centerline{Department of Astronomy, University of Michigan,
Ann Arbor, MI~~48109}
\centerline{and}
\centerline{Institute of Astronomy, University of Cambridge,}
\centerline{Madingley Road, Cambridge~~CB3 0HA, England$^{1}$}
\centerline{I: ssm@mail.ast.cam.ac.uk}
\vskip .25in
\centerline{J{\tenrm AMES} M.\ S{\tenrm CHOMBERT}$^{2}$}
\centerline{Infrared Processing and Analysis Center}
\centerline{California Institute of Technology, Pasadena, California~~91125}
\centerline{I: js@ipac.caltech.edu}
\vskip .1in
\centerline{\tenrm AND}
\vskip .1in
\centerline{G{\tenrm REGORY} D.\ B{\tenrm OTHUN}$^{3}$}
\centerline{Physics Department, University of Oregon, Eugene, OR~~97403}
\centerline{I: nuts@moo.uoregon.edu}
\vskip 1.0in
\centerline{\sl Accepted for publication in {\it The Astronomical Journal}}
\vskip 0.5in
$^1$Current address

$^2$Present address: Astrophysics Division, Code SZ,
NASA HQ, Washington, D.C., 20546

$^3$Visiting Astronomer, Kitt Peak National Observatory, National Optical
Astronomy Observatories, which is operated by Associated Universities for
Research in Astronomy, Inc., under cooperative agreement with the National
Science Foundation.
\eject
\centerline{\bf ABSTRACT}
\medskip

We present $UBVI$ and H$\alpha$ images of a sample of Low Surface Brightness
(LSB) disk galaxies.  These galaxies are generally late types, if they can
be sensibly classified at all.  However, they are not dwarfs, being
intrinsically large and luminous.

The morphology of LSB galaxies is discussed
in terms of the physical interpretation of the Hubble sequence.
Galaxies with high contrast relative to the sky background are subject
to being more finely typed than those which appear merely as fuzzy blobs
on photographic plates.  This causes
the stages of the Hubble sequence to be nonlinear in the sense
that large morphological type distinctions are made between high surface
brightness spirals when only small physical differences exist, and small
morphological distinctions are made between low surface brightness
galaxies even when large physical differences exist.

Many LSB galaxies lack the old red disk conspicuous in higher surface
brightness spirals.  Their morphology is strikingly similar in all bands
from $U$ to $I$, suggesting farily homogeneous stellar populations lacking
a well developed giant branch.  These properties, together with their
very blue colors, suggest that LSB galaxies are relatively younger than
their high surface brightness counterparts.  A few of these LSB galaxies
appear to be very young ($\simlt 1$~Gyr), and as such may represent
local examples of protogalaxies.

\vskip 0.5in
\medskip
\centerline{1. INTRODUCTION}
\medskip

Morphological classification of galaxies represents the earliest form
of extragalactic astronomy.  Detector technology initially
allowed only galaxies
of high contrast with respect to the sky background to be catalogued
and classified.   Selection effects have the potential to be severe in
this case.  However, in the last few years,
a large number of low contrast or Low Surface Brightness (LSB) galaxies have
been discovered and cataloged (Schombert \& Bothun 1988;
Schombert \etal 1992; Impey \etal 1994).  In general, LSB galaxies
span the same range of physical parameters as galaxies which occupy
the conventional Hubble sequence; they are not exclusively low mass
dwarf galaxies.  Since LSB galaxies are defined as
having central surface brightnesses fainter than the darkest night
sky, an investigation of their morphological properties may reveal
if they form some kind of hidden Hubble sequence.  Previous work
has firmly established that the physical properties of
LSB galaxies are strikingly different from those of the high surface brightness
(HSB) spirals which define the Hubble sequence (McGaugh 1992).  These
differences may provide important clues to the physics underlying morphology.

Morphological classification of galaxies has traditionally been done by
visual inspection of galaxy images on $B$-band photographic plates.  This
intrinsically non-linear process is difficult to duplicate using
linear, digital CCDs.  As such, morphological classification may be
more difficult and less precise in the digital era (see discussion in
van~den~Bergh, Pierce, \& Tully 1990).  To assist in the morphological
classification of LSB galaxies, in this paper we
present images in $U$, $B$, $V$, $I$, and H$\alpha$
of 22 LSB galaxies.  This is not a complete sample in any sense,
but does represent the lowest surface brightnesses disk galaxies
that can be extracted
from diameter limited field surveys using visual inspection
of plate material (\eg the UGC, Nilson 1973;
and the POSS-II LSB list, Schombert \etal 1992).  This population
of galaxies is very different from ``normal'' HSB field spirals and
LSB dwarfs in clusters (\eg Impey, Bothun,
\& Malin 1988; Irwin \etal 1990; Bothun, Impey, \& Malin 1991).

In \S 2 we discuss the morphology
of this sample of LSB galaxies, note their similarity to the excess
population of faint blue galaxies, and investigate the implications
that the physical properties of these galaxies have for the
interpretation of the Hubble sequence.  We comment on the CCD images of
interesting individual LSB galaxies in \S 3, some of which
are potentially young galaxies.  Our results are briefly summarized
in \S 4.  Throughout the paper, all distance dependent quantities
assume $H_0 = 100h\>{\rm km}\,{\rm s}^{-1}{\rm Mpc}^{-1}$
and a Virgo infall velocity of $300 \>{\rm km}\,{\rm s}^{-1}$.

\bigskip
\centerline{2. MORPHOLOGY}
\medskip

\smallskip
\centerline{2.1 {\sl LSB Galaxies}}
\smallskip

In general, LSB galaxies are late type (Sc and later) spiral and
irregular galaxies.  The spiral pattern is often incipient or
fragmentary and usually faint and difficult
to trace (\eg F530--3, F558--1, F561--1, F568--1,
F568--6, F577--V1, UGC 1230, UGC 5709, UGC 6151, UGC 6614, and UGC 9024).
The low visibility of the spiral pattern and frequently irregular
or amorphous appearance of the disk often result in a
dwarf classification, though few of the galaxies discussed here
are actually faint enough to formally qualify as such ($M > -16$).
Galaxies which do include F415--3 and F611--1, but, for instance,  UGC~12695 is
twice the size of the Milky Way despite its dwarf irregular appearance.
Bulge components are faint or
totally undetectable in most cases (\eg F561--1, UGC 1230, and UGC 6151),
but in a significant subset the bulges are bright.  These bulges generally
have effective radii of 2-3 kpc and are of normal surface brightness.
If anything, there is a tendency for the LSB galaxies
with prominent bulges to have larger, lower surface brightness disks
than the more typical bulgeless LSB galaxies (\eg F530-3, F568-6, UGC 9024, and
especially UGC 6614).  Bars are very rare, at least in
this $B$-selected sample.  As bars may be dynamical tracers of a previous
tidal interaction, their paucity in this sample may be another indication
of isolation on small scales (Bothun \etal 1993).
There are sometimes lens components in the
surface brightness profiles, but it is more common for the profile
to be simply exponential with some noise.

The original Hubble sequence ended with type Sc (Hubble 1936).
This particular stage encompasses a very wide range of intrinsic
galaxy properties, including many LSB galaxies.  In fact, most of
the highest surface brightness disks which are known (\eg M101)
are Sc galaxies.  Still, it
became necessary to extend the Hubble sequence to ever later types
as progressively better plate material was examined.  Interestingly,
prior to LSB galaxies being so named, they were predominant members
of the new Hubble type, Sd.  An excellent nearby example of this is
provided by NGC 247.
The majority of LSB galaxies fall in the late type bins of
this classification scheme
(to the extent that they do at all), as might be expected from the
nature of the selection effects which act against them
(Allen \& Shu 1979; McGaugh 1995).

\vfill\eject
\smallskip
\centerline{2.2 {\sl Faint Blue Galaxies?}}
\smallskip

As the limits of observation have been pressed ever deeper,
a large excess in the numbers counts of galaxies at faint magnitudes
has been noted (Tyson 1988; Lilly, Cowie, \& Gardner 1991;
Colless \etal 1991).  McGaugh (1994a) and Ferguson \& McGaugh (1995)
argued that this could be at least in part due to surface brightness
selection effects which act preferentially against LSB galaxies locally.
Recently, Griffiths \etal (1994) and Glazebrook \etal (1994) have
resolved galaxies in the magnitude range of the excess with {\sl HST}.
They find that the excess is due to peculiar, irregular looking galaxies, many
of which do not fit into the traditional classification scheme.  In addition,
the data set of Wirth, Koo, \& Kron (1994) on the high redshift cluster
CL0016+16 reveals a significant
difference in light concentration index ratio as a function of $g-r$ color:
the blue galaxies tend be the ones with the lowest light concentration indices.
This is consistent with the hypothesis of Rakos \& Schombert (1994)
that the Butcher-Oemler (1984) effect is due to the infall and subsequent
disruption of late type, LSB galaxies.

This new morphological information provided by the HST observations
is consistent with the scenario suggested by McGaugh (1994a), namely,
that the $z=0$ population of LSB disk galaxies have global properties quite
similar to the faint blue galaxy population.
Indeed, the simulated {\sl HST\/} images of Ferguson \& McGaugh (1995)
predict an excess of edge-on, fuzzy, and irregular objects over what is
expected from models based on standard galaxy mixes lacking LSB galaxies.
Though the simulations are intentionally extreme, they bear a greater
morphological resemblance to actual {\sl HST\/} data than does the
standard no evolution model, indicating that the nearby objects discussed
here may be very similar to what has been considered an
enigmatic population of faint blue galaxies.  If the local population of
LSB galaxies do indeed correspond to that responsible for the excess counts
at faint magnitudes, then the redshift distribution
of the galaxies classified as peculiar by Glazebrook \etal (1994)
should be skewed towards somewhat lower redshift than the normal
populations in the same
magnitude range.  However, the degree of the skew depends on the
precise form of the bivariate distribution and the true slope of
the faint end of the luminosity function (see Ferguson \& McGaugh 1995).

\medskip
\centerline{2.3 {\sl The Hubble Sequence}}
\smallskip

The study of galaxies as physical
objects is often expressed as an effort to understand the Hubble sequence.
Being essentially a matter of appearance, as principally manifested by
arm texture and definition,
classification along the Hubble sequence of spirals is
affected by the surface brightness of a galaxy through the contrast of its
features relative to the background.  A strong contrast
facilitates perception of
morphological distinctions, particularly when imaging with a non-linear
detector.   As a consequence, one might suspect that
the Hubble sequence is also nonlinear in the sense that objects with
high surface brightness would be more finely typed than others, since
it is easier to notice differences in texture at high contrast.

An important original motivation for galaxy morphological typing was the
success of stellar spectral typing in establishing the basic physical
properties of stars.  Though it is well established that mass is the
dominant parameter determining a star's position along the main sequence,
consider the steps required to get to it.  Observed spectral type is related to
temperature by one nonlinear transformation (one stage in spectral type
covers a variable range in temperature, with, for example,
F stars covering a small range in temperature).  Then there is another
nonlinear transformation
from temperature to mass (O stars representing a very large linear mass
range).  These effects lead to an HR diagram which has obvious main
sequence discontinuities when plotted in terms of spectral types
despite a presumably smooth underlying mass function (Houk \& Cowley 1975).

For stars, at least, these effects are well understood.  Such cannot be said
of the Hubble sequence.
The identification of the physical properties underlying it remains elusive.
Quantifiable global properties are not well correlated with Hubble type
(Boroson 1981; Kennicutt 1981; Bothun 1982; Kent 1985) and objective
techniques seem
to indicate that all characteristics of an image play a role in classification
(Storri-Lombardi \etal 1992).  Clearly it is necessary to sort out any
nonlinearities in the classification scheme before such trends that
do exist can be interpreted physically.

A basic problem, however, is that morphological classification is an inherently
nonquantitative process.  Despite the many quantitative measurements that
can now be made for galaxies, morphological classification still persists
as a substitute therefor.  For instance, a fundamental
property of galaxies which can be
quantified is their luminosity profiles.  For disk galaxies, these
have the form
$$ \mu(r) = \mu_0 + 1.086 {r \over \alpha}, \nexteq $$
where $\mu_0$ is the central surface brightness and $\alpha$ is the
scale length of the disk.  These characterize the luminosity density and
the size of a galaxy.  If morphological classification is to reflect underlying
physics, then the least one should hope for is a positive correlation
between measured galaxy structure and morphological type.

One complication is that the observed central surface brightness $\mu_0$ needs
to be inclination corrected to obtain the true
face on value, $\mu_0^c$.  Usually this is done by assuming that the
disks are optically thin, so that only edge brightening occurs.
This is a dubious assumption, but we retain it for consistency with
other published data as it is usually a small correction compared to the
range of surface brightness considered here and hence makes no difference
to the results.  If anything, the assumption of no extinction is more
appropriate in these LSB systems which are relatively dust free (McGaugh
1994b).

A link between $\mu_0^c$ and underlying physics was first suggested by
Freeman (1970), who found that all spirals had $\mu_0^c = 21.65 \pm 0.3\;B\sb$.
If this were a true physical result, and not due to selection effects,
then it would
imply that the processes of galaxy formation conspired to always arrive at a
particular mass surface density.  However, as noted by Schombert \etal (1992),
the very existence of disk galaxies with $\mu_0^c$
many standard deviations from the Freeman (1970) result indicates that
the surface brightness distribution is not so sharply peaked.
The LSB systems under consideration here fall far from the Freeman value, with
typical inclination corrected central surface brightnesses of
$\mu_0^c \approx 23.8\;B\sb$ ($7 \sigma$ deviant).  Other investigations
(McGaugh 1993; de Jong \& van der Kruit 1994;
Sprayberry 1994; McGaugh \etal 1995), demonstrate that the space density
of LSB galaxies is in fact similar to that of Freeman disks
(the number of galaxies with $\mu_0^c \approx 23.8$ catalogued
by Schombert \etal 1992 is approximately the same as the number of galaxies
with $\mu_0^c \approx 21.65$).
Moreover, the global properties of disk galaxies (\eg color, rotation velocity,
profile shape, \HI content) seem to be largely independent of $\mu_0^c$
(McGaugh 1992).

Figure~1 shows the distribution of disk galaxies in the $(\mu_0^c,\alpha)$
plane.  Data are taken from Romanishin, Strom, \& Strom (1983),
van~der~Kruit (1987), and McGaugh \& Bothun (1994).  The data of
McGaugh \& Bothun (1994) are based on the images presented below.
Though some dwarf galaxies with spiral structure do exist
(Schombert \etal 1995), here we exclude intrinsically small
($\alpha < 1\,h^{-1}{\rm kpc}$) galaxies since we wish to discuss
only those disk galaxies which are comparable in size to the spirals
which define the Hubble sequence.  As can be seen
from Fig.~1, the surface brightness of galaxies
is not simply a matter of size or morphological type, as
argued by van~der~Kruit (1987).

Figure~1 is analogous to the HR diagram for stars in the sense that
it plots the most fundamental observable properties of the objects
under consideration.  (Obviously, luminosity could be substituted
for either surface brightness or scale length, which would simply amount
to a transformation to coordinates delineated by the dotted lines.)
Unlike stars, and contrary to the hope of the morphological approach,
galaxies do not distinguish themselves much by Hubble type in this
diagram (see also de Jong \& van der Kruit 1994).  The entire plane
is occupied below luminosity and surface brightness maxima, with no
obvious tendency to congregate into distinct branches as with stars
in the HR diagram.  Indeed, we should consider it fortunate that
one physical parameter is so dominant for stars that variations
in others do not hopelessly
muddle the main sequence.  That there appears to be no main sequence
for galaxies suggests that no single parameter dominates.  Two
would seem to be a minimum, and mass and density make a reasonable
pair for beginning to explain the distribution in Fig.~1 (McGaugh 1992;
Mo, McGaugh, \& Bothun 1994).

Though morphology fails to distinguish galaxies in this fundamental plane,
there is a tendency for galaxies of low surface brightness
to be classified as late types.  However, all Hubble types cover the same range
in size.  Things morphologically classified as dwarfs are not necessarily
small (see also Schneider \etal 1992).
Late types do tend to be less luminous owing to their lower
surface brightness at a given size.  Examples of very large, luminous LSB
disks do exist; these tend to be labeled as relatively early types
because of their prominent bulges and anemic spiral
structure but probably represent a unique and
distinct class of galaxies (Schombert \etal 1992; Sprayberry \etal 1994).

Note that disk absolute magnitudes generally do not exceed $M_B = -21$,
and that this limit is approached by LSB as well as HSB galaxies.  This
result probably has the most to do with the physical conditions which
are required to produce a spiral galaxy, perhaps indicating the maximum
baryonic mass which has had time to cool.  Similarly, the Freeman (1970)
central surface brightness appears to be the maximum edge
of the distribution, perhaps suggesting that denser protogalaxies
become ellipticals.
Another interesting feature in
Fig.~1 is the apparent envelope demarcated by a line running from
$\alpha = 3$~kpc at the faintest surface brightness plotted to $\alpha =
10$~kpc at the brightest.  Though some giant LSB galaxies exist, there
does seem to be a slight tendency for larger disks to be higher in surface
brightness.  This is suggestive of a bivariate distribution in which there
is a modest correlation between size and surface brightness, and which
has a sharp decline in the density of galaxies larger than the envelope.
This is analogous to the knee of the luminosity function, but is not orthogonal
to the surface brightness axis as usually assumed.  Put another way,
the size distribution becomes steeper for lower surface brightnesses,
but also extends to larger sizes.  This implies that the luminosity function
steepens towards lower surface brightnesses, though presumably with
a lower normalization at $L^*$.  Interestingly, such a luminosity function,
when combined with a flat luminosity function for HSB galaxies, would
have a shape similar to that required to match the observed galaxy
counts (Koo, Gronwall, \& Bruzual 1993).  Unfortunately, it is
not possible to say from the present data if this is the case of even
if there is a real trend of this sort (see also Ferguson \& McGaugh 1995).

Though the Hubble type is loosely related to surface brightness, it is
interesting to
note that the overlap between types is substantial.  This is especially
true for early type disks which predominantly occur at high surface
brightnesses around the Freeman (1970) value.  There is no distinction
between S0/Sa, Sb, and to a lesser extent, Sc, galaxies in Fig.~1.
This of course means that other characteristics, such as arm texture,
play a dominant role in classification.  Such details are much
less striking in LSB galaxies in spite of the frequency of spiral structure.
In fact, surface brightness should be considered as another dimension
in a more proper 2-dimensional galaxy classification systems, such as
the RDDO system initiated by van den Bergh (1976).

Being based largely on arm texture and the tightness of the wrapping
of spiral arms, it is perhaps not surprising that
the Hubble sequence does not distinguish $\mu_0^c$ and $\alpha$
which are the basic properties of disks.  However, disks are not
the only component of the luminosity profile, and another aspect which
enters the morphological classification is the central concentration of
the light, often quantified by the bulge to disk ratio
$B/D$.  Simien \& de Vaucouleurs (1986) give a detailed formula relating
morphological type to mean $B/D$.  However, there is a great deal of scatter
in $B/D$ at a given type, much of which is real (Boroson 1981; Bothun 1982;
Kent 1985).  Low surface brightness galaxies pose particular difficulties
in this regard because of their apparent bimodal distribution of bulge
sizes. Most LSB galaxies have $B/D < 0.1$, but a significant subset have
$B/D \sim 1$ with no obvious transition population.
The classification of large bulge LSB spirals by arm texture can
not be reconciled with that by $B/D$.  Looking only at the arms, these are
late types.  But judging by $B/D$, they are early types.

This may just be a further indication that the Hubble sequence is
less fundamental
to the nature of galaxies as physical objects than might be hoped.
The fact that early type spirals
cluster strongly around the Freeman (1970) value, while later types
cover a much larger area in this parameter suggests that
the high contrast of HSB galaxies allow for fine distinctions in
appearance to be made between them.
LSB galaxies are lumped into a few late type bins because it is difficult
to see anything about them other than that they are fuzzy blobs.
This is reminiscent of Messier's objects which when initially discovered
were all faint diffuse blobs, unclassifiable using the imaging
capabilities of the time.
Thus it appears that the Hubble sequence is indeed nonlinear, providing
detailed information over a relatively small portion of the parameter space
occupied by galaxies where the contrast is the best.  There is therefore
a tendency to infer big differences in type from small physical differences
when the contrast is high, and little or no difference in type despite
large physical differences when the contrast is poor.

This is amply
clear in the football shaped classification diagram of de Vaucouleurs (1959),
which places the most emphasis on early type spirals.  While this is an
accurate portrayal of the classification system, the {\it physical\/}
parameter space continues to expand towards later types, which
exhibit a wide range of properties.
LSB galaxies alone cover a wide range in size as well as
surface brightness, and also in color (McGaugh \& Bothun 1994; de~Blok,
van der Hulst, \& Bothun 1995), metallicity
(McGaugh 1994b), gas content, and star formation properties (van~der~Hulst
\etal
1987, McGaugh 1992; van~der~Hulst \etal 1993).
Considering that LSB galaxies are actually quite common,
it is important to understand them as physical objects if we are to
decipher the meaning of the Hubble sequence.

By way of analogy, consider the case of stellar populations.  When Baade (1944)
first resolved stars in the bulge of the Andromeda galaxy, he was able
to introduce the concept of stellar populations with the aid of the
HR diagram for stars in our own galaxy as an interpretive tool.
Even though only the brightest stars were resolved, knowledge of the
HR diagram allowed assignment of the red bulge stars to a population similar
to that of globular clusters, while the bright blue stars of the disk
belonged to a population similar to that of the solar neighborhood.
However, the reverse is not possible --- without {\it a priori\/} knowledge
of the HR diagram, one could never infer the existence of the main sequence
from Baade's data.  Although some attempts have been made (\eg Whitmore
1984), there remains no physically understood equivalent of the
HR diagram for galaxies.  The Hubble sequence fails to provide one.
%

Though it has not succeeded as a tool for understanding galaxies the
way the HR diagram has for stars, the Hubble sequence does have merit.
Despite the large amount of real scatter, there is a clear trend
of $B/D$ with type.  There are large regions of parameter space
which real galaxies
do not occupy (\eg there are no giant HSB galaxies, \cf Kent 1985;
Sprayberry \etal 1994; de Jong \& van der Kruit 1994.),
and though the scatter again is large, the boundaries do
seem to be delineated by type.  Given that there is always a great
deal of scatter in plots involving morphological type, these uninhabited
regions of parameter space and the scatter itself may be telling us more about
disk galaxy formation than any of the weak trends that do exist.

Since the Hubble sequence is essentially one of regularity,
varying from early type galaxies which are smooth in appearance to
irregular late types, perhaps the most important physical characteristic
is the star formation time scale (\cf Kennicutt 1983; Kennicutt,
Tamblyn, \& Congdon 1994).
Most of the star formation in early type galaxies occurred long ago,
so dynamical processes have had time to regularize the appearance of
stochastic variations in bright star forming sites which are still apparent
in late types.  These latter have their star formation histories
weighted more towards
current epochs (Gallagher, Hunter, \& Tutukov 1984; McGaugh \& Bothun 1994).
The variation can not be in the absolute amount of star formation, present
or past, as this allows for no variation within a given
type.  Perhaps it is better described by the time scale
$\tau \sim \dot M_o/M$,
the current rate of star formation relative to the total integrated
star formation.  This in some sense is the inverse of the evolutionary
rate, which is rapid for early types and very slow for late types.
This would naturally explain the morphology--density (Dressler 1980)
and surface brightness--density (Bothun \etal 1993; Mo, McGaugh, \&
Bothun 1994) relations, because galaxies are expected to
form later and evolve more slowly the more isolated their progenitors
(see Mo \etal 1994).

\bigskip
\centerline{3. IMAGES}
\medskip

To give some impression of the galaxies we are discussing, and
an idea of the diversity of morphology of LSB disks,
we present multicolor CCD images and discuss some interesting individuals.
The depth of the CCD images gives rather more
information than is available on discovery plates; this can lead to
rather different morphological classifications.  This is illustrated in
Table~1, which compares the types given by the UGC and those
determined from the CCD images on the system described by Sandage \&
Binggeli (1984) as employed by Schombert \etal (1992).  Clearly, there is
a large uncertainty in type at low surface brightnesses,
as classification requires some eyeball interpolation
of unseen structures.  The UGC classifications have been retained in Fig.~1
since the other data are also photographic, but it is clear that morphological
type {\it per se\/} is not very meaningful for LSB galaxies.

Surface photometry and colors are discussed by McGaugh \& Bothun (1994),
as are details of the broad band observations.  The H$\alpha$ observations,
described by McGaugh (1992), provide the targets for \HII region
spectroscopy (McGaugh 1994b).  Like the broad band images, many of the
H$\alpha$ images were obtained with the MDM\footnote {$^1$}{MDM Observatory is
operated by the University of Michigan, Dartmouth College,
and the Massachusetts Institute of Technology.} 1.3 m telescope.  Some
of the H$\alpha$ images
were obtained with the KPNO 2.1 m telescope with typically longer exposures.
Hence these are considerably deeper, a point which should be kept in mind when
comparing the images.  They are denoted by ``2.1 m'' in the figure captions.

All images are presented with north up and east to the left.  Unless otherwise
noted, each image is $2.4'$ on a side.  The pixel scale is $0.48''$/pixel
for the 1.3 m images, and was binned to $0.38''$/pixel for the 2.1 m
images.  Most, though not all, data were obtained
under photometric conditions.  Images are scaled to have the same contrast
relative to the sky so as to reveal morphology;
saturation of the grayscale often happens at quite low surface brightness
which varies with the surface brightness of the galaxy.

\medskip
\centerline{3.1 {\it NGC 7757}}
\smallskip

This HSB disk galaxy is included for comparison (Fig.~2).  An Sc spiral,
it is the sort of galaxy most of us probably consider ``typical.''
Perhaps contrary to appearances, the exposure time and the signal to noise
ratio in the sky is less than in the other images.
If displayed in such a way as to make the sky value (rather than the contrast
relative to the sky) appear similar, all detail would be lost to saturation.

The disk of this galaxy is consumed by star formation with \HII regions
tracing a prominent spiral pattern.  Despite the
much higher star formation rate per unit area,
this galaxy is not as blue as most of the LSB galaxies discussed here.
Note the change in morphology with filter as the disk becomes
progressively smoother in the red, indicative of an old disk population.

\medskip
\centerline{3.2 {\it F415--3}}
\smallskip

This LSB galaxy has a single prominent \HII region near its center
and amorphous very low surface brightness plumes extending away from the
ends of the major axis of the main body (Fig.~3).  The western plume has
several knots which are prominent in the blue filters but which are not
obviously \HII regions.  This may be due to weak H$\alpha$ emission, or a
velocity difference which takes the line out of the narrow H$\alpha$ bandpass.

This relatively small (exponential scale length $\alpha = 1.2\,h^{-1}{\rm
kpc}$)
galaxy has a similar appearance in all filters, suggesting a fairly homogeneous
population which has not segregated dynamically.  The $V-I$ color ($V-I =
0.71$)
is remarkably blue.  Note the lack of an old red disk.  This is a generic
property of LSB galaxies, which often lack the diffuse red disks
associated with the old disk component in HSB galaxies.  This suggests
a population with a giant branch which is underdeveloped
due to a late commencement of star formation and a low mean age (McGaugh \&
Bothun 1994).  As indicated by their low surface brightness and the sparse
sprinkling of \HII regions across the disks, both the past and current
star formation rate per unit area is low.

\medskip
\centerline{3.3 {\it F469--2}}
\smallskip

This rather large ($\alpha = 2.9\,h^{-1}{\rm kpc}$) quite LSB ($\mu_0^c =
24.44\,B\sb$) galaxy has nonetheless several prominent \HII regions embedded
in a chaotic looking disk (Fig.~4).  This galaxy is also quite blue
($U-B = -0.44$, $B-V = 0.43$, $V-I = 0.94$)
with a similar morphology in each filter, again suggesting a youthful
population.  If anything, the disk (as opposed to the spiral pattern)
is less prominent in $I$, contrary to the case in NGC~7757.

\medskip
\centerline{3.4 {\it F530--3}}
\smallskip

This galaxy looks fairly normal, with a bulge and a two arm spiral pattern
(Fig.~5).  Nonetheless, it is much lower surface brightness ($\mu_0^c =
23.85\sb$) than the
HSB galaxy to the southeast, which is saturated in these images.  This
interloper is probably at a different redshift, as it is unusual for
large ($\alpha = 3.7\,h^{-1}{\rm kpc}$ in this case) LSB galaxies to have
nearby neighbors (Bothun \etal 1993, Mo \etal 1994).

\medskip
\centerline{3.5 {\it F561--1}}
\smallskip

A typical size ($\alpha = 2.6\,h^{-1}{\rm kpc}$) disk, this LSB
galaxy has a strong, one arm spiral pattern (Fig.~7).  Such features
are not uncommon in LSB galaxies, suggesting that $m = 1$ spiral modes
are possible at low surface densities.  Rings and single arms
indicate the importance of surface density thresholds (Kennicutt 1989)
to the star forming properties of LSB disks (van~der~Hulst \etal 1993).

\medskip
\centerline{3.6 {\it F563--V1}}
\smallskip

Despite its dwarf morphology, F563--V1 is not particularly small
($\alpha = 1.9\,h^{-1} {\rm kpc}$) (Fig.~8).  It nonetheless appears to have a
rather homogeneous stellar population.  It is also one of the most metal
poor extragalactic objects known (McGaugh 1994b), again suggestive of a
young object.

\medskip
\centerline{3.7 {\it F611--1}}
\smallskip

This is an otherwise amorphous galaxy with a pair of embedded knots of
star formation (Fig.~10).  These do not dominate the light as is often
the case in star bursting blue compact galaxies (BCGs).
While BCGs presumably have LSB progenitors (Tyson \& Scalo 1988),
the galaxies being discussed here are generally to large and bright
to be the progenitor population (though see Taylor, Brinks, \&
Skillman 1993) which probably involves galaxies which are smaller
and perhaps even lower surface brightness when not actively forming
stars.  Nonetheless, F611--1 is similar to BCGs in that it
is also quite metal poor ($Z \simlt 0.1 Z\solar$).  The uniformly
distributed light is blue, with $U-B = -0.24$ and $B-V = 0.44$.  Though
not terribly small ($\alpha = 1.5\,h^{-1}{\rm kpc}$), it is quite low surface
brightness ($\mu_0^c = 24.5\sb$).

\medskip
\centerline{3.8 {\it UGC 1230}}
\smallskip

This normal sized ($\alpha = 3\,h^{-1}{\rm kpc}$) disk is extremely
blue, with $V-I = 0.56$.  The galactic redenning is large ($A_B = 0.4$,
Burstein \& Heiles 1984), so the color depends on the assumed redenning
law.  However, the observed $V-I = 0.72$ is itself remarkably blue
for a composite stellar system.  The giant branch in this galaxy must
be quite feebly populated, indicating a young mean age.

There is a clear, if diffuse, spiral pattern (Fig.~12).
The spiral arms are not well traced by \HII
regions as in NGC~7757 though there are a few scattered about.
The pair of knots to the south which are prominent
in the $I$-band are also \HII regions.  That faint nebular emission is present
in these red clusters indicates that star formation has continued in the
vicinity long enough for the the first stars formed in the current episode
of star formation to evolve to the red supergiant phase.

\medskip
\centerline{3.9 {\it UGC 12695}}
\smallskip

This galaxy has an odd structure, with thick spoke-like structures
rather than arms projecting out from the center (Fig.~15).  These features
are present in all filters, indicating a fairly homogeneous stellar population.
There are a number of bright \HII regions around the edges of the galaxy.
In those to the west, a clear trend of color is present in the sense that
the southern clusters are bluer than the northern ones, perhaps indicating
the propagation of star formation in this direction.  The proximity of a
very red and blue cluster in the east may indicate a similar situation.
Since these star clusters evolve on short (a few $\times 10^7$~yr) time
scales, and the galaxy as a whole is quite blue ($B-V = 0.37$), star formation
could have propagated across the entire galaxy rather recently.  The colors
are consistent with an age of only a few $\times 10^8$~yr (\cf Salzer \etal
1991).  The gas mass
fraction is quite large and the stellar mass to light ratio is low (McGaugh
1992), also consistent with youth.  Thus, UGC~12695 seems to be an
example of a large ($\alpha \sim 6\,h^{-1}{\rm kpc}$) disk which has only
recently formed.  As such it is as reasonable a candidate ``protogalaxy''
as any, but has the advantage of being close enough to study in detail.

\vfill\eject
\medskip
\centerline{3.10 {\it F568--6}}
\smallskip

The second example of a giant ($\alpha \approx 16\,h^{-1}{\rm kpc}$)
low surface brightness disk, F568--6 (Fig.~16) is also known as
Malin~2 (Bothun \etal 1990).  This is the only LSB galaxy (other than
the relatively HSB UGC~5709) known to contain \HII regions with metallicities
higher than $\sim 0.3 Z\solar$ (McGaugh 1994b).  A wide range of metallicities
are present, suggestive of a steep abundance gradient, though the paucity
of \HII regions makes this difficult to determine.

\medskip
\centerline{3.11 {\it UGC 6614}}
\smallskip

UGC~6614 (Fig.~17) is comparable in size to Malin~2 ($\alpha \approx
12\,h^{-1}{\rm kpc}$).  A number of other giant disks are also known
(Impey \& Bothun 1989, Knezek 1993, Sprayberry \etal 1994), but none
approach the prototype Malin~1 in terms of scale length ($\alpha \approx
55\,h^{-1}{\rm kpc}$; Bothun \etal 1987).  The abundance
determinations for the \HII regions in UGC~6614 are ambiguous (McGaugh 1994b),
but suggest that unlike the majority of LSB galaxies, it could be quite
metal rich.  Since only the giant LSB galaxies exhibit high abundances,
and also tend to have more prominent bulge components, the disks of these
galaxies may have been polluted by metal production in the bulges
(Koeppen \& Arimoto 1990).  The disk of U6614 is very low in surface
brightness and extends well out of the field of view.  As evident
from its red integrated colors, the observed light is dominated
almost everywhere by the bulge.

\medskip
\centerline{3.12 {\it F577--V1}}
\smallskip

F577--V1 is not well described as an exponential disk, but is comparable
in size to a galaxy with $\alpha \approx 3\,h^{-1}{\rm kpc}$.  It is one
of the few late type LSB galaxies with a bar (Fig.~18) though such
features may be common in early type LSB galaxies (de~Blok, private
communication).  Its outer light profile
is dominated by a very blue, actively star forming spiral/ring structure.

\medskip
\centerline{3.13 {\it UGC 9024}}
\smallskip

This galaxy has a very low surface brightness disk ($\mu_0^c = 24.71\sb$) and
a fairly normal looking bulge.  Its over all color is quite blue ($B-I =
1.18$),
as is that of the bulge itself ($B-I = 1.51$).
Some hint of spiral structure is apparent in the $B$ image, and is also
traced by \HII regions in the deep H$\alpha$ image
(Fig.~23).  The disk is rather large ($\alpha = 5.6\,h^{-1}{\rm kpc}$),
and the presence of the bulge suggests that this may be a transition
object between normal sized, bulgeless LSB galaxies and the giant
cousins of Malin~1.  This galaxy represents a kind of discoverly limit
presented by photographic plate technology as its disk is barely discernible
on the original plate material.

\medskip
\centerline{3.14 {\it A Local LSB Galaxy --- IC 1613}}
\smallskip

Though not part of the sample discussed by McGaugh \& Bothun (1994),
the local group galaxy IC~1613 provides a good
demonstration of how pixel scale, resolution, and detector
technology drive morphological classification.
Figure~24 shows two images, both taken in the $B$ band but at a
different scale.
The first image was taken with a re-imaging camera (Aldering and Bothun 1991).
The resolution of IC~1613 into
clumps is apparent and its globally low surface brightness is not easily
appreciated.  The next image was taking with the Parking Lot
Camera system (Bothun and Thompson 1988), and shows roughly what
IC~1613 would look like if placed at a distance of 10 Mpc.

By the standards of the sample presented in this paper,
IC~1613 qualifies quantitatively as a low surface brightness
galaxy. It has a central surface brightness and scale length
determined from these images of $\mu_0 = 23.59 \pm 0.15\;B\sb$ and
$\alpha = 260 \pm 23''$, consistent with the values
reported by Hodge \etal 1991.  For the inclinations $i =50^{\circ}$
given by Hodge (1978), $\mu_0^c = 23.88$.  At a distance of 725~kpc
(Freedman 1988), the physical scale length is 0.9~kpc and the
absolute magnitude is $M_B = -14.68$.  This is in good agreement with
the value given by Hodge (1978) of $M_B = -14.63$.  However, only light
within a couple of scale lengths is actually detected;
integration of the surface brightness profile would make the galaxy
$\sim 0.5$~mag.\ brighter.  The galaxy is probably tidally truncated
so that this is an overestimate, but it does illustrate the need to
perform deep surface photometry to recover all the light of LSB
galaxies (McGaugh 1994a; McGaugh \& Bothun 1994, de~Blok \etal 1995).

IC~1613 is smaller and less luminous than the other galaxies presented here.
Unlike most of the LSB galaxies detected in field surveys like that
of Schombert \etal (1992), it is a true dwarf.  Morphologically,
however, it is indistinguishable from more distant LSB galaxies when
viewed at a comparable resolution.  When imaged like this,
it is just another fuzzy, diffuse member of the LSB galaxy population.

\bigskip
\centerline{4. CONCLUSIONS}
\medskip

We have presented multicolor CCD images of a sample of low surface brightness
disk galaxies.  We summarize our results as the following:

\item{1)} LSB galaxies exhibit a wide range of morphologies, commensurate
with the large area they occupy in the size--surface brightness plane.

\smallskip
\item{2)} LSB galaxies are generally late types, and vice-versa.  The
classifications are not particularly meaningful though, as they fail to
distinguish any obvious physical characteristics, with the possible vague
exception of galaxy evolutionary rate.

\smallskip
\item{3)} Not all LSB galaxies are dwarfs, in spite of their morphological
similarity to this class as a whole. For example, UGC 12695 is twice the
size of the Milky Way despite its dwarf irregular appearance.

\smallskip
\item{4)} Hubble type is loosely related to mean surface brightness,
however the overlap between types is substantial. Galaxy classification is
strongly dependent on characteristics such as arm texture and bulge to
disk ratio and, thus, is only weakly related to basic properties of the
disk such as $\mu_0^c$ or $\alpha$.

\smallskip
\item{5)} LSB galaxies divide into two types with respect to B/D ratio,
most with B/D $<$ 0.1 and a small, but significant subset with B/D
$\approx$ 1. Large bulge LSB spirals have a contradictory appearance
with late-type spiral features, yet large B/D ratio.

The large physical differences between LSB galaxies are not well
represented by morphological classification schemes, which tend to assign
them only a few vaguely defined types.  This suggests that the Hubble
sequence is nonlinear in that galaxies with high contrast relative to the
sky background are subject to being more finely typed than those which
appear merely as fuzzy blobs on photographic plates.  Hence the
differences between early type (Sa, Sb, and Sc) spirals, though real and
seemingly large, are in fact small compared to the entire volume of
physical parameter space occupied by disk galaxies.

Many LSB galaxies are morphologically similar to the irregular faint blue
galaxies resolved by {\sl HST\/}.  They typically lack the old red disk
conspicuous in higher surface brightness spirals.  Their appearance does
not vary significantly with band from $U$ to $I$, suggesting fairly
homogeneous stellar populations.  Together with their blue colors, this
suggests that LSB galaxies are relatively young galaxies.

\vfil\eject

\bigskip
\centerline{\bf REFERENCES}
\bigskip
\frenchspacing

\ref Aldering, G.~S., \& Bothun, G.~D. 1991, PASP, 103, 1296

\ref Allen, R.~J., \& Shu, F.~H. 1979, ApJ, 227, 67


\ref Baade, W. 1944, ApJ, 100, 137

\ref Boroson, T. 1981, ApJS, 46, 177


\ref Bothun, G.~D. 1982, ApJS, 50, 39



\ref Bothun, G.~D., Impey, C.~D., \& Malin, D.~F. 1991, ApJ, 376, 404

\ref Bothun, G.~D., Impey, C.~D., Malin, D.~F., \& Mould, J.~R.
	1987, AJ, 94, 23



\ref Bothun, G.~D., Schombert, J.~M., Impey, C.~D., \& Schneider, S.~E.
	1990, ApJ, 360, 427

\ref Bothun, G.~D., Schombert, J.~M., Impey, C.~D.,
	Sprayberry, D., \& McGaugh, S.~S. 1993, AJ, 106, 548

\ref Bothun, G.~D., \& Thompson, I.~B. 1988, AJ, 96, 877

\ref Burstein, D., \& Heiles, C. 1984, ApJS, 54, 33

\ref Butcher, H., \& Oemler, A. 1984, ApJ, 285, 426

\ref Colless, M., Ellis, R.~S., Taylor, K. \& Shaw, G. 1991
        MNRAS, 253, 686


\ref de Blok, W.~J.~G., van~der~Hulst, J.~M., \& Bothun, G.D.
	1995, submitted



\ref de Jong, R.~S., \& van der Kruit, P.~C. 1994, A\&AS, 106, 451

\ref de Vaucoulers, G. 1959, in {Handbuch der Physik, v.~53,
       Astrophysics~IV:  Stellar Systems}, ed. S. Fl\"ugge (Berlin:
       Springer-Verlag), 275




\ref Dressler, A.  1980, ApJ, 236, 351



\ref Ferguson, H.~C., \& McGaugh, S.~S. 1995, ApJ, in press


\ref Freedman, W.~L. 1988, ApJ, 326, 691

\ref Freeman, K.~C. 1970, ApJ, 160, 811

\ref Gallagher, J.~S.~I., Hunter, D.~A., \& Tutukov, A.~V. 1984,
	ApJ, 284, 544

\ref Glazebrook K., Ellis R. S., Santiago B. X., Elson R. A. W.,
	Griffiths R. E., \& Ratnatunga K. 1994, preprint

\ref Griffiths, R. E., Casertano, S., Ratnatunga, K. U., Neuschaefer,
	L. W., Ellis, R. S., Gilmore, G. F., Glazebrook, K., Santiago, B.,
	Huchra, J. P., Windhorst, R. A., Pascarelle, S. M., Green, R. F.,
	Illingworth, G. D., Koo, D. C., \& Tyson, J. A. 1994, ApJ, in press

\ref Hodge, P.~W. 1978, ApJS, 37, 145

\ref Hodge, P.~W., Smith, T.~R., Eskridge, P.~B., MacGillivray, H.~T.,
	\& Beard, S.~M. 1991, ApJ, 369, 372

\ref Houk, N., \& Cowley, A.~P. 1975, {University of Michigan Catalogue
	of Two-Dimension\-al Spectral Types for the HD Stars}, Vol. 1

\ref Hubble, E.~P. 1936, {The Realm of the Nebulae} (New Haven:
        Yale University Press)




\ref Impey, C.~D., \& Bothun, G.~D. 1989, ApJ, 341, 89

\ref Impey, C.~D., Bothun, G.~D., \& Malin, D.~F. 1988, ApJ, 330, 634

\ref Impey, C.~D., Sprayberry, D., Bothun, G.~D., \& Irwin, M.~J.
	1994, BAAS, 25, 1291

\ref Irwin, M.~J., Davies, J.~I., Disney, M.~J., \& Phillipps, S.
	1990, MNRAS, 245, 289




\ref Kennicutt, R.~C. 1981, AJ, 86, 1847

\ref Kennicutt, R.~C. 1983, ApJ, 267, 563

\ref Kennicutt, R.~C. 1989, ApJ, 344, 685

\ref Kennicutt, R.~C., Tamblyn, P., \& Congdon, C.~W. 1994, ApJ, 435, 22

\ref Kent, S. 1985, ApJS, 59, 115

\ref Knezek, P. 1993, Ph.D. Thesis, University of Massachusetts

\ref Koeppen, J., \& Arimoto, N. 1990, A\&A, 240, 22

\ref Koo, D.~C., Gronwall, C., \& Bruzual, A.~G. 1993, ApJ, 415, L21


\ref Lilly, S.~J., Cowie, L.~L., \& Gardner, J.~P. 1991, ApJ, 369, 79


\ref McGaugh, S.~S. 1992, Ph.D. Thesis, University of Michigan

\ref McGaugh, S.~S. 1993, BAAS, 25, 1398

\ref McGaugh, S.~S. 1994a, Nature, 367, 538

\ref McGaugh, S.~S. 1994b, ApJ, 426, 135

\ref McGaugh, S.~S. 1995, in preparation

\ref McGaugh, S.~S., \& Bothun, G.~D. 1994, AJ, 107, 530


\ref McGaugh, S.~S., Bothun, G.~D., Schombert, J.~M. 1995, submitted

\ref Mo, H.~J., McGaugh, S.~S., \& Bothun, G.~D. 1994, MNRAS, 267, 129





\ref Rakos, K.~D, \& Schombert, J.~M. 1994, ApJ, in press

\ref Romanishin, W., Strom, K.~M., \& Strom, S.~E. 1983, ApJS, 53, 105

\ref Salzer, J.~J., Alighieri, S.~D.~S., Matteucci, F.,
	Giovanelli, R., \& Haynes, M.~P. 1991, AJ, 101, 1258

\ref Sandage, A., \& Binggeli, B. 1984, AJ, 89, 919

\ref Schneider, S.~E., Thuan, T.~X., Mangum, J.~G., \& Miller, J.
	1992, ApJS, 81, 5


\ref Schombert, J.~M., \& Bothun, G.~D. 1988, AJ, 95, 1389


\ref Schombert, J.~M., Bothun, G.~D., Schneider, S.~E., \&
	McGaugh, S.~S. 1992, AJ, 103, 1107

\ref Schombert, J.~M., Pildis, R.~A., Eder, J., \& Oemler, A.
	1995, in preparation

\ref Simien, F., \& de Vaucouleurs, G. 1986, ApJ, 302, 564


\ref Sprayberry, D. 1994, Ph.D. thesis, University of Arizona

\ref Sprayberry, D., Impey, C.~D., Irwin, M., McMahon, R.~G., \&
        Bothun, G.~D. 1994, ApJ, in press

\ref Storrie-Lombardi, M.~C., Lahav, O., Sodre, L., \& Storrie-Lombardi,
	L.~J. 1992, MNRAS, 259, 8P

\ref Taylor, C., Brinks, E., \& Skillman, E.~D. 1993, AJ, 105, 1


\ref Tyson, J.~A. 1988, AJ, 96, 1

\ref Tyson, N.~D., \& Scalo, J.~M. 1988, ApJ, 329, 618


\ref van~den~Bergh, S. 1976, ApJ, 206, 883

\ref van den Bergh, S., Pierce, M.~J., \& Tully, R.~B. 1990, ApJ, 359, 4

\ref van~der~Hulst, J.~M., Skillman, E.~D., Kennicutt, R.~C.,
        \& Bothun, G.~D. 1987, A\&A, 177, 63

\ref van der Hulst, J.~M., Skillman, E.~D., Smith, T.~R.,
        Bothun, G.~D., McGaugh, S.~S., \& de~Blok, W.~J.~G. 1993, AJ, 106, 548

\ref van~der~Kruit, P.~C. 1987, A\&A, 173, 59

\ref Whitmore, B.~C. 1984, ApJ, 278, 61

\ref Wirth, G.~D., Koo, D.~C., and Kron, R.~G. 1994, preprint

\nonfrenchspacing
\vfil\eject
\bigskip
\centerline{\bf FIGURE CAPTIONS}
\bigskip
\item{\bf Figure 1.} The distribution of disk galaxies in the central
surface brightness -- exponential scale length plane.  Different Hubble types
are distinguished with different symbols.  The dotted lines are lines of
constant luminosity labeled by absolute $B$ magnitude.  The solid
and dashed horizontal lines represent the Freeman (1970) result
$\mu_0^c = 21.65 \pm 0.3~B\sb$.  Note that disks of all sizes exist with
surface brightnesses many $\sigma$ below the Freeman (1970) value, and that
there is no obvious discontinuity in either surface brightness or size
over a range of four magnitudes.  Galaxies inhabit the entire ($\mu_0,\alpha$)
plane below maxima in each.  The Hubble sequence fails to delineate any
meaningful relations between these fundamental parameters.
\item{\bf Figure 2.} NGC 7757. a) $U$ b) $B$ c) $V$ d) $I$
e) H$\alpha$ f) continuum subtracted H$\alpha$.  This HSB spiral is
included for comparison purposes.  The grayscale is not scaled the same
as for the following LSB images; if it were the entire disk would be
saturated.
\item{\bf Figure 3.} F415--3. a) $U$ b) $B$ c) $V$ d) $I$
e) H$\alpha$ f) continuum subtracted H$\alpha$.
\item{\bf Figure 4.} F469--2. a) $U$ b) $B$ c) $V$ d) $I$
e) H$\alpha$ f) continuum subtracted H$\alpha$.
The linear streaks result
from deferred charge from bright stars in previous exposures.
\item{\bf Figure 5.} F530--3. a) $U$ b) $B$ c) $V$ d) $I$
e) H$\alpha$ f) continuum subtracted H$\alpha$.
\item{\bf Figure 6.} F558--1. a) $U$ b) $B$ c) $V$ d) $I$
e) 2.1 m H$\alpha$ f) continuum subtracted H$\alpha$.  Since the H$\alpha$
images were obtained with a larger telescope than for the previous galaxies
and filters, they are rather deeper (see text).
A fairly normal looking spiral, the broad band images were unfortunately
not obtained under photometric conditions so it is difficult to usefully
employ the color information.  Despite the imperfect continuum subtraction,
there are \HII regions confirmed by spectroscopy in the southern arm.
\item{\bf Figure 7.} F561--1. a) $U$ b) $B$ c) $V$ d) $I$
e) 2.1 m H$\alpha$ f) continuum subtracted H$\alpha$.
\item{\bf Figure 8.} F563--V1. a) $U$ b) $B$ c) $V$ d) $I$
e) 2.1 m H$\alpha$ f) continuum subtracted H$\alpha$.
The $U$ band image is not photometric,
and this galaxy is rather bluer than seems to be indicated by the
low contrast in (a) (de~Blok \etal 1995).
\item{\bf Figure 9.} F563--V2. a) $U$ b) $B$ c) $V$ d) $I$
e) H$\alpha$ f) continuum subtracted H$\alpha$.
This galaxy has two \HII regions
west of the relatively HSB central bar, and a blue, LSB plume to the
northwest.
\item{\bf Figure 10.} F611--1. a) $U$ b) $B$ c) $V$ d) $I$
e) H$\alpha$ f) continuum subtracted H$\alpha$.
\item{\bf Figure 11.} F746--1. a) $U$ b) $B$ c) $V$ d) $I$
e) H$\alpha$ f) continuum subtracted H$\alpha$.
There are a number of star forming regions in this relatively
HSB galaxy.
\item{\bf Figure 12.} UGC 1230. a) $U$ b) $B$ c) $V$ d) $I$
e) H$\alpha$ f) continuum subtracted H$\alpha$.  The various knots
are \HII regions spanning a wide range of intrinsic H$\alpha$ luminosity
and broad band color.  For example,
compare the blue northeastern knots to the red ones due south of them.
Though the H$\alpha$ emission of the southern knots is barely discernible
in this image, it was easily detected spectroscopically (McGaugh 1992, 1994b).
\item{\bf Figure 13.} UGC 5709. a) $U$ b) $B$ c) $V$ d) $I$
e) 2.1 m H$\alpha$ f) continuum subtracted H$\alpha$.
This galaxy is intermediate in surface brightness, and displays
the same sort of old red disk seen in HSB galaxies.  It is
also intermediate in color, being redder than the lower surface
brightness disks (McGaugh \& Bothun 1994).
\item{\bf Figure 14.} UGC 6151. a) $U$ b) $B$ c) $V$ d) $I$
e) 2.1 m H$\alpha$ f) continuum subtracted H$\alpha$.
There are quite a few faint \HII regions in this irregular
LSB spiral.
\item{\bf Figure 15.} UGC 12695. a) $U$ b) $B$ c) $V$ d) $I$
e) H$\alpha$ f) continuum subtracted H$\alpha$.  Aside from the radial
spokes (rather than spiral arms), this galaxy presents a rather
chaotic appearance suggestive of recent coalescence.  The colors
of the \HII knots imply rapid evolution and propagation of star
formation (see text).  The two brightest \HII regions each have
H$\alpha$ luminosities requiring ionization by $\sim 10,000$ O stars.
\item{\bf Figure 16.} F568--6 $=$ Malin~2. a) $U$ b) $B$ c) $V$ d) $I$
e) 2.1 m H$\alpha$.  An off band image adequate for continuum
subtraction is not available.  The $B$ and $I$ images were obtained
on a different observing run than the $U$ and $V$ images, the latter
being nonphotometric.  However, it is useful to intercompare the emission
regions visible in the $U$.  The \HII regions are predictably bright in
this filter, but the oblong structure northwest of the bulge is not.
The spectrum of this object is consistent with shock heating (McGaugh 1994b),
so it may be a jet associated with the nuclear activity in this giant
galaxy (Bothun \etal 1990).
\item{\bf Figure 17.} UGC 6614. a) $U$ b) $B$ c) $V$ d) $I$
e) 2.1 m H$\alpha$.  An off band image adequate for continuum
subtraction is not available.  The scale of these images is different
from the others, being 3.1 rather than 2.4 arcminutes on a side.
Spiral structure can be traced to the edge of the frame, and
extends well beyond this.
\item{\bf Figure 18.} F577--V1. a) $U$ b) $B$ c) $V$ d) $I$.
\item{\bf Figure 19.} F568--1. a) $B$ b) $I$ c) 2.1 m H$\alpha$
d) continuum subtracted H$\alpha$.
\item{\bf Figure 20.} F583--5. a) $B$ b) $I$ c) 2.1 m H$\alpha$
d) continuum subtracted H$\alpha$.
\item{\bf Figure 21.} F585--3. a) $B$ b) $I$ c) 2.1 m H$\alpha$
d) continuum subtracted H$\alpha$.
\item{\bf Figure 22.} UGC 5675. a) $B$ b) $I$ c) 2.1 m H$\alpha$
d) continuum subtracted H$\alpha$.
\item{\bf Figure 23.} UGC 9024. a) $B$ b) $I$ c) 2.1 m H$\alpha$
d) continuum subtracted H$\alpha$.
\item{\bf Figure 24.} IC~1613.  Both images are in the $B$ band.
Because IC~1613 is in the local group, these CCD images had to be obtained
with reducing optics (see text).
a) This image is $15'$ on a side with a pixel scale of $3.05''$/pixel.
The brighter stars are clearly resolved.
b) This image is $51'$ on a side with a pixel scale of $10.2''$/pixel.
This is 3.4 times larger than (a), and shows what IC~1613 would look
like if unresolved --- a typical LSB galaxy.

\bye